\documentclass[sigconf]{acmart}

\setcopyright{none}

\acmConference[FAccTRec '21]{FAccTRec '21: The 4th FAccTRec Workshop on Responsible Recommendation}{September 25, 2021}{Amsterdam, Netherlands}

\begin{document}

\title{Recommendation Fairness: From Static to Dynamic}

\author{Dell Zhang}
\authornote{[DISCLAIMER] The opinions expressed herein are our own and do not necessarily represent the views of our employers.}
\authornote{Dell Zhang is on leave from Birkbeck, University of London, and working full-time for ByteDance AI Lab.}
\email{dell.z@ieee.org}
\orcid{0000-0002-8774-3725}
\affiliation{%
  \institution{ByteDance AI Lab}
  \city{London}
  \country{UK}
}

\author{Jun Wang}
\authornotemark[1]
\email{jun.wang@cs.ucl.ac.uk}
\affiliation{%
  \institution{University College London}
  \city{London}
  \country{UK}
}



\begin{abstract}
Driven by the need to capture users' evolving interests and optimize their long-term experiences, more and more recommender systems have started to model recommendation as a Markov decision process and employ \emph{reinforcement learning} to address the problem.
Shouldn't research on the fairness of recommender systems follow the same trend from static evaluation and one-shot intervention to dynamic monitoring and non-stop control? 
In this paper, we portray the recent developments in recommender systems first and then discuss how fairness could be baked into the reinforcement learning techniques for recommendation. 
Moreover, we argue that in order to make further progress in recommendation fairness, we may want to consider multi-agent (game-theoretic) optimization, multi-objective (Pareto) optimization, and simulation-based optimization, in the general framework of \emph{stochastic games}.
\end{abstract}

\maketitle

\section{Introduction}

Fairness in recommendation is one of the most important aspects of evaluating and providing socially responsible recommender systems.
In this short position paper, we briefly review the research about recommendation fairness over the last five years and present our opinions about where this area should go next.
The central idea is that the paradigm of recommendation fairness will be shifting from static evaluation and one-shot intervention to dynamic monitoring and non-stop control. 

\section{The Rise of Reinforcement Learning for Recommendation}

The mainstream approach to building recommender systems is to formulate recommendation as \emph{matrix completion}~\cite{jannachRecommenderSystemsMatrix2016}, i.e., given a matrix of users by items where the value at cell $(i,j)$ represents the $i$-th user's rating of the $j$-th item, predict the missing cell values based on the existing cell values.
Such a matrix completion problem is often solved by \emph{matrix factorization}~\cite{korenMatrixFactorizationTechniques2009,rendleFactorizationMachines2010} algorithms. 
Recently, the nonlinear neural variants of matrix factorization~\cite{covingtonDeepNeuralNetworks2016,heNeuralCollaborativeFiltering2017,heNeuralFactorizationMachines2017} for recommendation have attracted a lot of attention, though some researchers have their reservations~\cite{ludewigPerformanceComparisonNeural2019,rendleNeuralCollaborativeFiltering2020}.

Today, modern recommender systems underpinning various web or mobile apps are expected to become more personalized and interactive so as to better serve the users. 
Consequently, traditional recommendation techniques based on matrix completion which assume users' preferences being static and aim to maximize their immediate satisfaction would no longer work well.

\emph{Reinforcement learning} (RL)~\cite{suttonReinforcementLearningIntroduction2018} --- an area of machine learning which is concerned with optimal decision making over time in a dynamic environment --- offers a promising approach to tackling the problems of personalization and interactivity by capturing users' evolving interests and optimizing their long-term experiences~\cite{zhaoDeepReinforcementLearning2019}. 
Inspired by the great successes of reinforcement learning, particularly when it is combined with \emph{deep learning}~\cite{lecunDeepLearning2015,goodfellowDeepLearning2016} such as in AlphaGo~\cite{silverMasteringGameGo2016}, reinforcement learning based recommender systems~\cite{kuhnleReinforcementLearningInformation2021,zhangDeepReinforcementLearning2020,zhangDRL4IR2ndWorkshop2021} have just started to gain popularity in the last couple of years.

\subsection{Recommendation as an MDP}

To apply reinforcement learning to recommender systems, we need to model recommendation as a \emph{Markov decision process} (MDP)~\cite{suttonReinforcementLearningIntroduction2018}. 
For example, in a video recommender system~\cite{chenTopKOffPolicyCorrection2019,maOffpolicyLearningTwostage2020}, the state space $\mathcal{S}$ describes the users each accompanied with their contextual status (e.g., the time when the recommendation is made and the query text entered by the user), and the action space $\mathcal{A}$ consists of all possible video items available for recommendation.
The state representing each user evolves as the user interacts with the recommender system; different users will have different states.
Using reinforcement learning, we seek a policy $\pi(a|s)$ which returns a distribution of video items $a \in \mathcal{A}$ for each given user state $s \in \mathcal{S}$.
The objective is that the learned policy $\pi$ can maximize the expected discounted cumulative reward over potentially infinite time horizon, where the immediate reward for taking action $a$ at state $s$ (i.e., recommending item $a$ to user $s$) is defined by a reward function $R:\mathcal{S}\times\mathcal{A} \to \mathbb{R}$.
 
\subsection{Reinforcement Learning Algorithms in Recommender Systems}

Many different reinforcement learning algorithms have been employed by a variety of recommender systems in recent years. 
The simplest ones~\cite{glowackaBanditAlgorithmsInformation2019,barraza-urbinaIntroductionBanditsRecommender2020} utilize contextual (multi-armed) \emph{bandits}~\cite{slivkinsIntroductionMultiArmedBandits2019,lattimoreBanditAlgorithms2020} which solve the special case of one-step reinforcement learning;
some~\cite{zhengDRNDeepReinforcement2018,zhaoRecommendationsNegativeFeedback2018,zhaoJointlyLearningRecommend2020} use value-based methods such as the Deep Q Network (DQN)~\cite{mnihHumanLevelControlDeep2015,hesselRainbowCombiningImprovements2018};
some~\cite{wangIRGANMinimaxGame2017,chenTopKOffPolicyCorrection2019,maOffpolicyLearningTwostage2020} use policy-based methods such as the policy gradient algorithm REINFORCE~\cite{suttonReinforcementLearningIntroduction2018,williamsSimpleStatisticalGradientFollowing1992}; and 
some~\cite{xinSelfSupervisedReinforcementLearning2020,chenReinforcementRecommendationUser2021} adapt actor-critic methods~\cite{kondaActorCriticAlgorithms1999,silverDeterministicPolicyGradient2014}.
A hot research topic is to develop \emph{offline reinforcement learning}~\cite{levineOfflineReinforcementLearning2020,kidambiMOReLModelBasedOffline2020,yangRepresentationMattersOffline2021} methods for interactive recommendation~\cite{zouPseudoDynaQReinforcement2020,xiaoGeneralOfflineReinforcement2021} which can make effective use of previously collected user-item interaction data without expensive online data collection. 

\section{The Evolution of Recommendation Fairness}

Significant progress has been made in the space of \emph{algorithmic fairness}~\cite{saleiroDealingBiasFairness2020,mehrabiSurveyBiasFairness2021} for recommender systems~\cite{ekstrandFairnessDiscriminationRecommendation2019,gaoCounteractingBiasIncreasing2020,liTutorialFairnessMachine2021}, as reflected by the FAccTRec workshop series since it first took place in 2017~\cite{ekstrandFATRECWorkshopResponsible2017,kamishima2ndFATRECWorkshop2018,ekstrand3rdFAccTRecWorkshop2020}. 

\subsection{Static Recommendation Fairness}

It has been discovered and reported that various types of bias could exist in recommender systems, such as those with respect to user demographics (gender, age, and race, etc.)~\cite{yaoParityFairnessObjectives2017,zhuFairnessAwareTensorBasedRecommendation2018,geyikFairnessAwareRankingSearch2019}, user activeness~\cite{fuFairnessAwareExplainableRecommendation2020}, and item popularity~\cite{abdollahpouriControllingPopularityBias2017,abdollahpouriUnfairnessPopularityBias2019,abdollahpouriConnectionPopularityBias2020,zhangCausalInterventionLeveraging2021}. 
Hence, many different metrics of fairness~\cite{castelnovoZooFairnessMetrics2021} have been proposed in order to build fairness-aware recommender systems. 
Such recommendation fairness metrics can be defined at two levels: individual fairness and group fairness.
Generally speaking, the techniques for counteracting bias and promoting fairness in recommendation so far are largely in the form of constrained optimization~\cite{zafarFairnessConstraintsFlexible2019}: either maximizing utility (which is for the most part the relevance of recommended items to users) subject to a set of fairness constraints~\cite{celisControllingPolarizationPersonalization2019,geyikFairnessAwareRankingSearch2019}, or maximizing fairness subject to a lower bound of utility~\cite{zhuFairnessAwareTensorBasedRecommendation2018}, or jointly optimizing both for an overall satisfaction~\cite{gaoHowFairCan2019}. 
Those works mostly make fairness adjustments to traditional matrix completion based recommender systems.
Their concept of recommendation fairness is static in the sense that the protected groups are fixed during the recommendation process.

\subsection{Dynamic Recommendation Fairness}

Little research has been conducted to investigate the fairness of reinforcement learning based recommender systems where the protected groups may change over time.  
In dynamic environments where the distribution of population is shifting or the decisions being made have feedback effects, counter-intuitive phenomena (like the Simpson's paradox) could occur and biases could be iteratively amplified. 
For example, imposing static fairness criteria myopically at every step may actually exacerbate unfairness~\cite{liuDelayedImpactFair2018,creagerCausalModelingFairness2020,damourFairnessNotStatic2020,mouzannarFairDecisionMaking2019,williamsDynamicModelingEquilibria2019,zhangHowFairDecisions2020}.

Note that although there exist a few papers talking about fairness in reinforcement learning in general, not all those fairness definitions are related to the fair treatment of different users or items (grouped by sensitive \emph{protected} attributes) in recommendation. 
For example, the ``meritocratic'' fairness of reinforcement learning~\cite{jabbariFairnessReinforcementLearning2017} does not seem to be very relevant to our fairness concern for recommender systems.

Liu et al.~\cite{liuBalancingAccuracyFairness2021} have proposed a fairness-aware recommendation framework based on reinforcement learning to dynamically balance recommendation accuracy and user fairness in the long run. 
The constantly changing user preferences and fairness statuses are jointly represented as the states in the MDP model of recommendation.  
Furthermore, a two-fold reward function is designed to combine accuracy and fairness.
Also on the user-side of fairness but not particularly for recommender systems, Wen et al.~\cite{wenAlgorithmsFairnessSequential2021} studied reinforcement learning under group fairness constraints. 
They show how fairness constraints from the supervised learning setting such as demographic parity and equality of opportunity can be extended to the MDP setting.
The algorithms developed by them to solve the MDP problem can ensure that the learned policy does not favor the majority sub-population over the minority sub-population.
Zhang et al.~\cite{zhangHowFairDecisions2020} investigated specifically the dynamics of group qualification rates~\cite{mouzannarFairDecisionMaking2019,williamsDynamicModelingEquilibria2019} under the more general partially-observed MDP setting.
Moreover, there also exist some studies of fairness-aware (contextual) multi-armed bandits~~\cite{barraza-urbinaIntroductionBanditsRecommender2020,glowackaBanditAlgorithmsInteractive2017,glowackaBanditAlgorithmsInformation2019,slivkinsIntroductionMultiArmedBandits2019,lattimoreBanditAlgorithms2020} --- a simplified form of reinforcement learning --- where the fairness constraint is defined as a minimum rate at which a task/resource is assigned to a user~\cite{chenFairContextualMultiArmed2020,patilAchievingFairnessStochastic2020,claureMultiArmedBanditsFairness2020,liCombinatorialSleepingBandits2019,xuCombinatorialMultiArmedBandits2020}.

Ge et al.~\cite{geLongTermFairnessRecommendation2021} have made an attempt on dynamic recommendation fairness of not users but items.
Their work focuses on the the fairness to different groups of items divided according to their degrees of popularity which dynamically change during the recommendation process: popular items could become unpopular after a while and vice versa.
To achieve long-term fairness in terms of item exposure, the MDP model of recommendation is augmented with a set of fairness constraints each of which is an auxiliary fairness cost function bounded by the corresponding limit.
Such a constrained MDP problem can be solved by performing constrained policy optimization with an actor-critic architecture.

When a recommender system starts to utilize reinforcement learning to optimize its users' long-term engagement, there will be a risk of the unethical phenomenon ``user tampering''~\cite{evansUserTamperingReinforcement2021,carrollEstimatingPenalizingPreference2021} whereby the recommender system tries to actively manipulate its users' preferences via its recommendations in order to gain maximum accumulated reward.
For example, a news recommender system may be tempted to (politically) polarize its users with the early recommendations so that the users will become more engaged with the system's later recommendations catering to such polarization.
Obviously, fairness issues will arise if different users are affected by user tampering differently.
How to ensure dynamic recommendation fairness while addressing user tampering is still an open question.

\subsection{Looking Ahead}

\subsubsection{Multi-Agent (Game-Theoretic) Optimization}

Since recommender systems are by their nature multi-sided platforms or marketplaces~\cite{singhFairnessExposureRankings2018,morikControllingFairnessBias2020} involving at least the consumers (customers) of items as well as the producers (providers) of items~\cite{surerMultistakeholderRecommendationProvider2018,burkeRecommendationMultistakeholderEnvironments2019}, there have been some works on optimizing the static fairness for all stakeholders of the recommender system~\cite{burkeBalancedNeighborhoodsMultisided2018,mehrotraFairMarketplaceCounterfactual2018,suhrTwoSidedFairnessRepeated2019,patroFairRecTwoSidedFairness2020,mansouryFairnessAwareRecommendationMultiSided2021}.
The usual approach to multi-sided static fairness is to use a linear interpolation of all stakeholders' fairness metrics~\cite{suhrTwoSidedFairnessRepeated2019,patroFairRecTwoSidedFairness2020,wuTFROMTwosidedFairnessAware2021} as the optimization objective or constraint. 
That is probably not sophisticated enough to handle the intricacies of dynamic fairness in reinforcement learning.

We think that the principled approach to multi-sided dynamic fairness for recommender systems is to consider it as a \emph{multi-agent reinforcement learning} (MARL)~\cite{yangOverviewMultiAgentReinforcement2021} problem, as in \cite{zhangFairnessMultiAgentSequential2014,jiangLearningFairnessMultiAgent2019,zimmerLearningFairPolicies2021}. 
This implies adopting the mathematical framework of \emph{stochastic games}~\cite{littmanMarkovGamesFramework1994,leyton-brownEssentialsGameTheory2008,shohamMultiagentSystemsAlgorithmic2008} which generalize MDPs to multiple interacting decision makers.
Note that to maximize the total social welfare while ensuring each agent (player) get a fair share of opportunities, we may want to go beyond \emph{Nash equilibrium} and embrace the more general \emph{correlated equilibrium} which is also computationally more efficient. 
The advantages of correlated equilibrium over Nash equilibrium in terms of fairness could be illustrated by the classic ``battle of the sexes'' game~\cite{leyton-brownEssentialsGameTheory2008,shohamMultiagentSystemsAlgorithmic2008}. 
Furthermore, \emph{behavior game theory} may be useful for analyzing multi-sided dynamic fairness in real-life recommender systems.
As revealed by the observation of human playing the ``ultimatum bargaining'' game, people have the tendency to pursue fairness even when it contradicts with the subgame perfect equilibrium that maximizes their monetary payoffs~\cite{leyton-brownEssentialsGameTheory2008,shohamMultiagentSystemsAlgorithmic2008}.


\subsubsection{Multi-Objective (Pareto) Optimization}

It is probably also time to leave the constrained optimization~\cite{zafarFairnessConstraintsFlexible2019,gaoHowFairCan2019} approach to recommendation fairness behind and seek the Pareto optimization~\cite{chenReinforcementRecommendationUser2021} of multiple objectives including the utility and the fairness.
Even when fairness is the only goal of our concern, researchers have rigorously proved that there are inherent conflicts among some common fairness metrics and it is often impossible to optimize them simultaneously~\cite{kleinbergInherentTradeOffsFair2016,friedlerImPossibilityFairness2021}.
A few early works have emerged~\cite{xiaoFairnessAwareGroupRecommendation2017,siddiqueLearningFairPolicies2020} and more should follow the steps.

\subsubsection{Simulation Environment}

To fully comprehend and tackle the complexities of dynamic fairness in recommendation, it is highly desirable to develop a simulation environment for such multi-agent multi-objective recommender systems where a number of fairness metrics could be continuously monitored and optimized.  
Although Google has released \texttt{Fairness-Gym}~\cite{damourFairnessNotStatic2020} for the simulation of simple dynamic fairness tasks (loan application~\cite{liuDelayedImpactFair2018}, college admission~\cite{huDisparateEffectsStrategic2019,milliSocialCostStrategic2019}, and attention allocation~\cite{ensignRunawayFeedbackLoops2018,elzaynFairAlgorithmsLearning2019}), a simulator dedicated to dynamic fairness in recommendation is not available, yet.
This would require more fundamental research on understanding user behavior and building user models in recommender systems~\cite{puUserCentricEvaluationFramework2011,balogTransparentScrutableExplainable2019,ekstrandSimuRecWorkshopSynthetic2021}.
It also looks promising to incorporate fairness metrics and models into some newly emerging probabilistic simulators of multi-agent recommender systems such as \texttt{RecSim NG}~\cite{mladenovRecSimNGPrincipled2021}.

\section{Conclusion}

The recent developments in the area of recommendation fairness exhibits a clear trend towards the dynamic view of fairness.
Accordingly, the underlying mathematical framework of fair recommendation will probably move from matrix completion to Markov decision process and then to stochastic games.
Such new models and algorithms for fairness may not only improve different kinds of recommender systems but also have impacts upon the broader field of Responsible AI.

\begin{acks}
We are grateful to Dr Hang Li, Director of AI Lab at ByteDance, for his guidance on and support for this line of research. 
We thank the workshop organizers for providing this forum and the reviewers for their constructive feedback.
\end{acks}

\bibliographystyle{ACM-Reference-Format}
\bibliography{ref_fairness,ref_explainability,ref_information-retrieval,ref_reinforcement-learning,ref_game-theory}


\end{document}